\documentstyle[twoside,fleqn,espcrc2]{article}

\newcommand{\AmS}{{\protect\the\textfont2
  A\kern-.1667em\lower.5ex\hbox{M}\kern-.125emS}}

\newcommand{\be}{\begin{equation}}                                              
\newcommand{\ee}{\end{equation}}                                                
\newcommand{\half}{\frac{1}{2}}

\title{SUSY on the lattice}

\author{I. Montvay
\address{Deutsches Elektronen Synchrotron DESY, \\
         Notkestr.~85, D-22603 Hamburg, Germany}}
       
\begin{document}

\begin{abstract}
 The motivation and perspectives of numerical simulations of
 supersymmetric Yang-Mills theories are reviewed.
\end{abstract}

\maketitle

%%%%%%%%%%%%%%%%%%%%%%%%%%%%%%%%%%%%%%%%%%%%%%%%%%%%%%%%%%%%%%%%%%%%%%%%
\section{INTRODUCTION}                                      \label{sec1}

 Supersymmetric quantum field theories, reviewed for instance in refs.
 \cite{WESBAG}-\cite{SOHNIUS}, have peculiar properties.
 They represent an interesting special subset of quantum field theories.
 Because of their highly symmetric nature, they are best suited for
 analytical studies, which sometimes lead to exact solutions
 \cite{SEIWIT}.

%%%%%%%%%%%%%%%%%%%%%%%%%%%%%%%%%%%%%%%%%%%%%%%%%%%%%%%%%%%%%%%%%%%%%%%%
\subsection{Supersymmetric Yang-Mills theory}             \label{sec1.1}

 Since local gauge symmetries play a very important r\^ole in Nature,
 there is a particular interest in supersymmetric gauge theories.
 The simplest examples are supersymmetric Yang-Mills (SYM) theories
 which are supersymmetric extensions of pure gauge theories.
 Depending on the number $N$ of supersymmetry generator pairs
 $Q_{i\alpha}, \bar{Q}_{i\dot{\alpha}}$,  ($i=1,2,\ldots,N$), one speaks
 about $N=1$ ``simple'' supersymmetry or $N \geq 2$ ``extended''
 supersymmetry.
 Interesting examples of SYM theories have, for instance, $N=1$ or 2
 supersymmetry.

 The SYM action with $N=1$ supersymmetry can be written as follows:
\begin{eqnarray}\nonumber
S_{SYM} & = & \frac{1}{4\pi}\, {\rm Im} \left\{ \tau
\int d^4x\, d^2\theta\, {\rm Tr}(W^\alpha W_\alpha) \right\}
\\ 
\nonumber
 & = & \frac{1}{g^2}\int d^4x\, {\rm Tr}\left[ -\half
 F_{\mu\nu}F^{\mu\nu}
\right.
\\ 
\nonumber
 & & \left. -\; i\lambda\sigma^\mu(D_\mu\bar{\lambda})
+i(D_\mu\bar{\lambda})\bar{\sigma}^\mu\lambda + D^2
\right]
\\
\label{eq01} 
 & & +\; \frac{\Theta}{16\pi^2} \int d^4x\, 
{\rm Tr} \left[ F_{\mu\nu}\tilde{F}^{\mu\nu} \right] \ .
\end{eqnarray}
 Here $\tau$ is a complex coupling parameter
\be\label{eq02}
\tau \equiv \frac{\Theta}{2\pi} + \frac{4\pi i}{g^2}
\ee
 combining the gauge coupling $g$ with the $\Theta$-parameter.
 The first line of eq.~(\ref{eq01}) is in superfield notation, in terms
 of the spinorial field strength superfield
 $W(x,\theta,\bar{\theta})_\alpha$ which depends on the four-coordinate
 $x$ and the anticommuting Weyl-spinor variables
 $\theta_\alpha,\bar{\theta}_{\dot\alpha}$ ($\alpha,\dot{\alpha}=1,2$).
 After performing the grassmanian integration on $\theta$, one obtains
 the second form of the SYM action in eq.~(\ref{eq01}) in terms of the
 component fields: the field strength tensor $F_{\mu\nu}$, its dual
 $\tilde{F}_{\mu\nu} \equiv \half\epsilon_{\mu\nu\rho\sigma}
 F_{\rho\sigma}$, the Majorana fermion field in the adjoint
 representation $\lambda,\bar{\lambda}$ and the auxiliary field $D$.
 Of course, $D_\mu$ denotes gauge covariant derivative and
 $\sigma^\mu \equiv \{1,\sigma^{1,2,3}\}$.
 (For details of the formalism see refs.~\cite{WESBAG}-\cite{SOHNIUS}.)

 Since the auxiliary field without kinetic term $D$ can be trivially
 integrated out, $S_{SYM}$ is nothing else but a Yang-Mills theory
 with a massless Majorana fermion in the adjoint representation.
 Hence the Yang-Mills theory of such a field, called ``gaugino'' or, in
 the context of strong interactions, ``gluino'', is automatically
 supersymmetric.
 Introducing a non-zero gaugino mass $m_{\tilde{g}}$ breaks
 supersymmetry ``softly''.
 Such a mass term is:
\be\label{eq03}
m_{\tilde{g}} (\lambda^\alpha\lambda_\alpha + 
\bar{\lambda}^{\dot{\alpha}} \bar{\lambda}_{\dot{\alpha}} )
= m_{\tilde{g}} (\bar{\Psi}\Psi) \ .
\ee
 Here in the first form the Majorana-Weyl components
 $\lambda,\bar{\lambda}$ are used, in the second form the Dirac-Majorana
 field $\Psi$.

 $S_{SYM}$ in eq.~(\ref{eq01}) defines the simplest, $N=1$ SYM theory.
 It is reasonable to start the non-perturbative lattice studies by this.
 An important feature of this theory is its global $U(1)_\lambda$ chiral
 symmetry which coincides with the so called \linebreak
 {\em R-symmetry} generated by the transformations
\be\label{eq04}
\theta_\alpha^\prime = e^{i\varphi}\theta_\alpha \ , \hspace{2em}
\bar{\theta}_{\dot{\alpha}}^\prime = 
e^{-i\varphi}\bar{\theta}_{\dot{\alpha}} \ .
\ee
 This is equivalent to
\be\label{eq05}
\lambda_\alpha^\prime = e^{i\varphi}\lambda_\alpha \ , \hspace{0.5em}
\bar{\lambda}_{\dot{\alpha}}^\prime = 
e^{-i\varphi}\bar{\lambda}_{\dot{\alpha}} \ , \hspace{0.5em}
\Psi^\prime = e^{-i\varphi\gamma_5}\Psi \ .
\ee

 The $U(1)_\lambda$-symmetry is anomalous: for the corresponding axial
 current $J_\mu \equiv \bar{\Psi}\gamma_\mu\gamma_5\Psi$, in case of
 $SU(N_c)$  gauge group with coupling $g$, we have
\be\label{eq06}
\partial^\mu J_\mu = \frac{N_c g^2}{32\pi^2} 
\epsilon^{\mu\nu\rho\sigma} F_{\mu\nu}^r F_{\rho\sigma}^r \ .
\ee
 However, the anomaly leaves a $Z_{2N_c}$ subgroup of $U(1)_\lambda$
 unbroken.
 This can be seen, for instance, by noting that the transformations
\be\label{eq07}
\Psi \to e^{-i\varphi\gamma_5}\Psi \ , \hspace{2em}
\bar{\Psi} \to \bar{\Psi}e^{-i\varphi\gamma_5}
\ee
 are equivalent to
\be\label{eq08}
m_{\tilde{g}} \to m_{\tilde{g}} e^{-2i\varphi\gamma_5} \ ,\;
\Theta_{SYM} \to \Theta_{SYM} - 2N_c\varphi \ ,
\ee
 where $\Theta_{SYM}$ is the $\Theta$-parameter of gauge dynamics.
 Since $\Theta_{SYM}$ is periodic with period $2\pi$, for
 $m_{\tilde{g}}=0$ the $U(1)_\lambda$ symmetry is unbroken if
\be\label{eq09}
\varphi = \varphi_k \equiv \frac{k\pi}{N_c} \ , \hspace{2em}
(k=0,1,\ldots,2N_c-1) \ .
\ee
 For this statement it is essential that the topological charge is
 integer.

 The discrete global chiral symmetry $Z_{2N_c}$ is expected to be 
 spontaneously broken by the non-zero {\em gaugino condensate}
 $\langle \lambda\lambda \rangle \ne 0$ to $Z_2$ defined by 
 $\{\varphi_0,\varphi_{N_c}\}$ ($\lambda \to -\lambda$ is a rotation).
 Instanton calculations (for a review and references see \cite{AKMRV})
 give at $\Theta_{SYM}=0$ in $N_c$ degenerate vacua ($k=0,\ldots,N_c-1$)
\be\label{eq10}
\frac{g^2}{32\pi^2}\langle \lambda\lambda \rangle_0
= \frac{4\,e^{2\pi ik/N_c}}{[(N_c-1)!(3N_c-1)]^{1/N_c}}
\Lambda_{SYM}^3
\ee
 Here $\Lambda_{SYM}$ denotes the $\Lambda$-parameter of the gauge
 coupling.
 The transformation with $\Theta_{SYM}$ is
\be\label{eq11}
\langle \lambda\lambda \rangle_{\Theta_{SYM}} = e^{i\Theta_{SYM}/N_c}
\langle \lambda\lambda \rangle_0 \ .
\ee

 In general, an important and interesting question is whether
 supersymmetry can be broken spontaneously or not.
 In pure SYM theory, without additional matter supermultiplets, this is
 not expected to occur.
 An argument for this is given by the non-zero value of the {\em Witten
 index} \cite{WITTEN}
\be\label{eq12}
w \equiv {\rm Tr\,}(-1)^F = n_{boson}-n_{fermion} \ ,
\ee
 which is equal to the difference of the number of bosonic minus
 fermionic states with zero energy.
 It is supposed not to change with the parameters of the theory.
 For SYM theory we have $w_{SYM}=N_c$, therefore there is no spontaneous
 supersymmetry breaking.
 (The $N_c$ ground states discussed above correspond to this Witten
 index.)
\vspace*{-0.7em}

%%%%%%%%%%%%%%%%%%%%%%%%%%%%%%%%%%%%%%%%%%%%%%%%%%%%%%%%%%%%%%%%%%%%%%%%
\subsection{Seiberg-Witten solution}                      \label{sec1.2}

 The SYM theory with $N=2$ extended supersymmetry is a highly
 constrained theory which has, however, more structure than the
 relatively simple $N=1$ case discussed above.
 In particular, besides the $N=1$ ``vector superfield'' containing the
 gauge boson and gaugino $(A_\mu,\lambda)$, it also involves an $N=1$
 ``chiral superfield'' $(\phi,\lambda^\prime)$ in the adjoint
 representation which consists of the complex scalar $\phi$ and the
 Majorana fermion $\lambda^\prime$.
 The Majorana pair $(\lambda,\lambda^\prime)$ can be combined to a
 Dirac-fermion $\psi$ and then the vector-like (non-chiral) nature of
 this theory can be made explicit.
 
 The main new feature of $N=2$ SYM is that it contains also scalar
 fields, hence there is the possibility of Higgs mechanism.
 Let us here consider only an $SU(2)$ gauge group, when the Higgs
 mechanism implies the symmetry breaking $SU(2) \to U(1)$.
 The complex expectation value of the scalar field
 $\langle \phi \rangle$ parametrizes the {\em moduli space} of
 zero-energy degenerate vacua.
 The degeneracy is a consequence of extended supersymmetry.
 Due to the Higgs mechanism the ``charged'' gauge bosons become heavy
 and the low-energy effective theory is $N=2$ SYM with $U(1)$ gauge
 group.

 As it has been shown by Seiberg and Witten \cite{SEIWIT}, extended 
 supersymmetry and asymptotic freedom can be used to determine exactly
 the low-energy effective action in the Higgs phase, if the vacuum
 expectation value is large.
 It turns out that, besides the singularity which corresponds to
 asymptotic freedom at infinity, at strong couplings there are two
 singularities of the effective action which describe light monopoles
 and dyons, respectively.

 The unprecedented achievement of an exact solution for a non-trivial
 four-dimensional quantum field theory induced a considerable activity
 among theoreticians.
 Recently there is a new way of considering non-perturbative problems
 in quantum field theory.
 It is an interesting question how much the more conventional approach
 based on lattice regularization can contribute to these new insights.

%%%%%%%%%%%%%%%%%%%%%%%%%%%%%%%%%%%%%%%%%%%%%%%%%%%%%%%%%%%%%%%%%%%%%%%%
\section{LATTICE FORMULATION}                               \label{sec2}

 The simplest supersymmetric gauge theory is $N=1\; SU(2)$ SYM.
 Massive gluinos (Majorana fermions in the adjoint representation)
 break supersymmetry softly.
 As discussed in section~\ref{sec1.1}, the supersymmetric point is at
 $m_{\tilde{g}}=0$.

%%%%%%%%%%%%%%%%%%%%%%%%%%%%%%%%%%%%%%%%%%%%%%%%%%%%%%%%%%%%%%%%%%%%%%%%
\subsection{Actions}                                      \label{sec2.1}

 The {\em Curci-Veneziano action} of $N=1$ SYM \cite{CURVEN} is based on
 Wilson fermions.
 The effective gauge action obtained after integrating out the gluino
 field is given by
\be\label{eq13}
S_{CV} = \beta\sum_{pl} \left( 1-\half{\rm Tr\,}U_{pl} \right)
- \half\log\det Q[U] \ ,
\ee
 where the fermion matrix is
\be\label{eq14}
Q_{yv,xu} = \delta_{yx}\delta_{vu} - 
K\sum_{\mu=\pm} \delta_{y,x+\hat{\mu}}(1+\gamma_\mu) V_{vu,x\mu} 
\ee
 and the gauge link in the adjoint representation is defined as
\be\label{eq15}
V_{rs,x\mu} = \half
{\rm\,Tr\,}(U_{x\mu}^\dagger \tau_r U_{x\mu} \tau_s ) \ .
\ee

 The Majorana nature of the gaugino is represented in eq.~(\ref{eq13})
 by the factor $\half$ in front of $\log\det Q$.
 There is no problem with the sign of $\det Q$ because we have
 $\det Q=\det\tilde{Q}$, where
\be\label{eq16}
\tilde{Q} \equiv \gamma_5 Q = \tilde{Q}^\dagger
\ee
 is the hermitean fermion matrix.
 From the relations
\be\label{eq17}
CQC^{-1} = Q^T \ , \hspace{1em} B\tilde{Q}B^{-1} = \tilde{Q}^T \ ,
\ee
 with the charge conjugation matrix $C$ and $B \equiv C\gamma_5$,
 it follows that every eigenvalue of $Q$ and $\tilde{Q}$ is (at least)
 doubly degenerate.
 Therefore, with the (real) eigenvalues $\tilde{\lambda}_i$ of
 $\tilde{Q}$, we have
\be\label{eq18}
\det\tilde{Q} = \prod_i \tilde{\lambda}_i^2 \geq 0 .
\ee
 The doubling of the eigenvalues of $\tilde{Q}$ and/or $Q$ implies that
 for fermions in the adjoint representation zero eigenvalues always
 occur in pairs, as in the continuum \cite{HSU}.
 (This is different for fermions in the fundamental representation:
 see for instance \cite{NARVRA}.)

 Of course, the lattice action in eq.~(\ref{eq13}) is not unique.
 Another possibility is based on five-dimensional domain walls
 \cite{DOMAIN}.
 In this approach one knows the value of the bare fermion mass where
 the supersymmetric continuum limit is best approached and one may have
 advantages from the point of view of the speed of symmetry restoration,
 but one has to pay with the proliferation of (auxiliary) fermion
 flavours.
 There are also speculations about direct dimensional reduction on the
 lattice from $d=10$ or $d=6$ dimensions down to the physically
 interesting case $d=4$, started by ref.~\cite{HUNANE} and continued by
 ref.~\cite{MARNISH}.
 However, up to now it has not yet been demonstrated that this approach
 does really work. 

%%%%%%%%%%%%%%%%%%%%%%%%%%%%%%%%%%%%%%%%%%%%%%%%%%%%%%%%%%%%%%%%%%%%%%%%
\subsection{Algorithms}                                   \label{sec2.2}

 As one can see in the lattice action in eq.~(\ref{eq13}), Majorana
 fermions effectively mean a flavour number $N_f=\half$.
 This can be achieved by the {\em hybrid classical dynamics algorithm}
 \cite{GLTRS}, which is applicable to any number of flavours.
 Since it is a finite step size algorithm, an extrapolation to zero
 step size is needed.
 This approach has been implemented and successfully tested for the
 Curci-Veneziano action (\ref{eq13}) by Donini and Guagnelli in
 ref.~\cite{DONGUA}.

 Another possibility to simulate non-even numbers of flavours is based
 on the {\em multi-bosonic algorithm} \cite{LUSCHER}.
 A two-step variant using a {\em noisy correction step} \cite{KENKUT}
 has been developed in ref.~\cite{GLUINO}.
 This algorithm is based, for general $N_f$, on the approximation
\be\label{eq19}
\det(Q^\dagger Q)^{N_f/2} \simeq \frac{1}{\det P_n(Q^\dagger Q)} \ ,
\ee
 where the polynomial $P_n$ satisfies
\be\label{eq20}
\lim_{n \to \infty} P_n(x) = x^{-N_f/2}
\ee
 within a suitably chosen positive interval $x \in [\epsilon,\lambda]$,
 which covers the spectrum of $Q^\dagger Q$ in typical gauge
 configurations.
 In the two-step variant one uses a first polynomial $\bar{P}_{\bar{n}}$
 for a crude approximation and realizes a fine correction by another
 polynomial $P_n$, where now
\be\label{eq21}
\lim_{n \to \infty} \bar{P}_{\bar{n}}(x)P_n(x) = 
x^{-N_f/2} \;\;{\rm for}\;\; x \in [\epsilon,\lambda] \ .
\ee
 The fermion determinant is approximated as
\be\label{eq22}
\det(Q^\dagger Q)^{N_f/2} \simeq
\frac{1}{\det \bar{P}_{\bar{n}}(Q^\dagger Q) \det P_n(Q^\dagger Q)} \ .
\ee
 The advantage of this two-step variant is its smaller storage
 requirement and shorter autocorrelations, because the order of the
 first polynomial $\bar{n}$ can be kept small.
 (In praxis, for instance, $\bar{n} \leq 10-12$ for lattice sizes
 $\leq 8^3\cdot 16$.)
 As a consequence of the smallness of $\bar{n}$, this remains a large
 step size algorithm, also in the limit of good precision.
 
 In order to achieve an exact algorithm in this two-step variant of the
 multi-bosonic algorithm, in principle, an extrapolation to
 $n \to \infty$ is needed, together with measurement-corrections for
 eigenvalues outside $[\epsilon,\lambda]$.
 In praxis, however, $(\bar{n},n)$ and $[\epsilon,\lambda]$ can be
 chosen such that these kinds of corrections are smaller than the
 statistical errors.

\begin{figure}[htb]
\vspace*{7.0cm}
\includegraphics{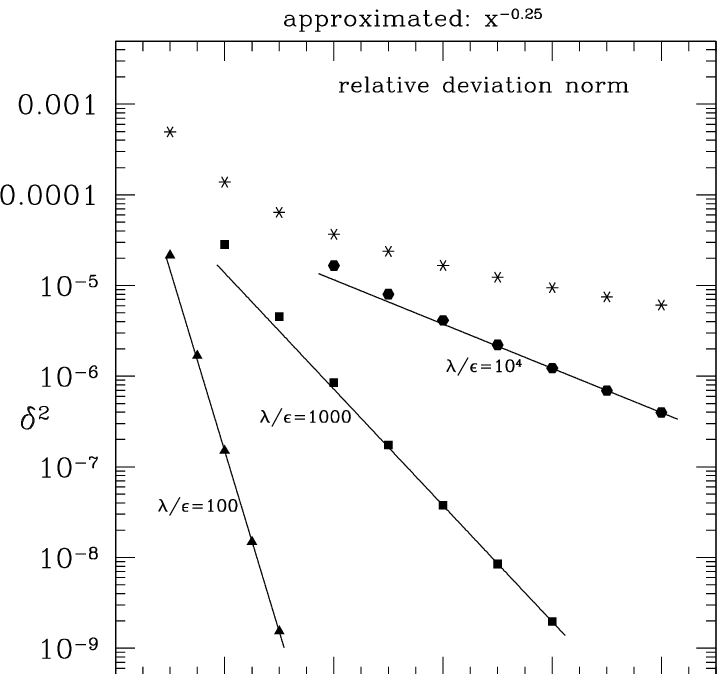}
\caption{The deviation norm $\delta^2$ of the polynomial approximations
 of $x^{-1/4}$ as function of the order for different values of
 $\lambda/\epsilon$.
 The asterisks show the $\epsilon/\lambda \to 0$ limit.
 (Ref.~\cite{POLYNOM})}
\vspace*{-0.7cm}
\label{fig01}
\end{figure}

 Random matrix models suggest that the fluctuations of the minimal
 eigenvalue are inversely proportional to the lattice volume.
 (For references and a recent summary see ref.~\cite{VERBAAR}.)
 This is advantageous for the choice of the interval
 $[\epsilon,\lambda]$.
 For the calculation of the necessary polynomials procedures written
 in Maple are available \cite{POLYNOM}.
 The quality of the polynomial approximations is illustrated by
 figure~\ref{fig01}.
%%%%%%%%%%%%%%%%%%%%%%%%%%%%%%%%%%%%%%%%%%%%%%%%%%%%%%%%%%%%%%%%%%%%%%%%
%%%%%%%%%%%%%%%%%%%%%%%%%%%%%%%%%%%%%%%%%%%%%%%%%%%%%%%%%%%%%%%%%%%%%%%%

 Besides the above ``hermitean'' version ``complex'' (non-hermitean)
 variants of the multi-bo\-so\-nic algorithms \cite{BOFOGA} can also be
 exploited, preferably also in two-step variants.
 In order to improve performance, preconditioning according to
 ref.~\cite{JEGERL} turned out to be very useful.

%%%%%%%%%%%%%%%%%%%%%%%%%%%%%%%%%%%%%%%%%%%%%%%%%%%%%%%%%%%%%%%%%%%%%%%%
\subsection{Quenched computations}                        \label{sec2.3}

\begin{figure}[htb]
\vspace*{5.9cm}
\includegraphics{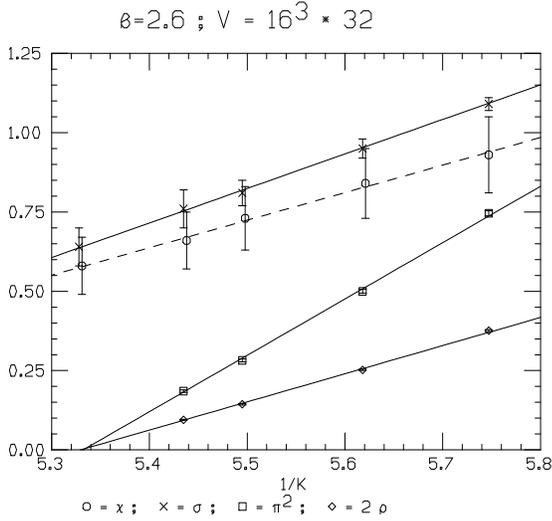}
\caption{Masses of $\tilde{\pi},\;\tilde{\sigma},\;\tilde{\chi}$ and
 $2\rho$ as a function of $1/K$ (the mass of $\tilde{\pi}$ is squared).
 The extrapolation at $K_{cr}$ is also shown (curves). 
 (Ref.~\cite{DOGUHEVL})}
\vspace*{-0.7cm}
\label{fig02}
\end{figure}

 Quenched calculations without the effect of dynamical gluinos can be
 useful for roughly localizing the critical region in the hopping
 parameter $K$ and for testing the methods to determine masses.
 Such calculations have been performed recently in
 refs.~\cite{KOUMON} and \cite{DOGUHEVL}.
 An example of the obtained lowest masses is shown in 
 figure~\ref{fig02}.
 Notation conventions are as follows: $\tilde{\pi}$ and $\tilde{\sigma}$
 denote, respectively, the pion-like and sigma-like states made out of
 gluinos.
 $\tilde{\chi}$ is a Majorana fermion state made out of gluons and 
 gluinos. 
%%%%%%%%%%%%%%%%%%%%%%%%%%%%%%%%%%%%%%%%%%%%%%%%%%%%%%%%%%%%%%%%%%%%%%%%
%%%%%%%%%%%%%%%%%%%%%%%%%%%%%%%%%%%%%%%%%%%%%%%%%%%%%%%%%%%%%%%%%%%%%%%%

%%%%%%%%%%%%%%%%%%%%%%%%%%%%%%%%%%%%%%%%%%%%%%%%%%%%%%%%%%%%%%%%%%%%%%%%
\section{FIRST RESULTS AND OUTLOOK}                         \label{sec3}

 The goal of the DESY-M\"unster-Athens collaboration \cite{DEMUAT} is to
 study  non-perturbative properties of SYM theories by Monte Carlo
 simulations.
 The computations are performed on the APE-Quadrics at DESY-Zeuthen and
 on the CRAY-T3E at HLRZ-J\"ulich by using the two-step multi-bosonic
 fermion algorithm for gluinos discussed in section~\ref{sec2.2}.

%%%%%%%%%%%%%%%%%%%%%%%%%%%%%%%%%%%%%%%%%%%%%%%%%%%%%%%%%%%%%%%%%%%%%%%%
\subsection{Chiral symmetry breaking}                     \label{sec3.1}

 An important first step in the numerical investigations is to find
 the critical hopping parameter $K_{cr}$ corresponding to zero gluino
 mass.
 As it is discussed in section~\ref{sec1.1}, the supersymmetric point
 is at zero gluino mass.
 This tells how the supersymmetric continuum limit $\beta \to \infty$
 has to be performed in the $(\beta,K)$-plane.
 The basic expectation is that at $K=K_{cr}$, for gauge group
 $SU(N_c)$, there are $N_c$ degenerate ground states with different
 values of the gluino condensate (see eq.~(\ref{eq10})).
 This corresponds to the expected spontaneous global chiral symmetry
 breaking pattern $Z_{2N_c} \to Z_2$ \cite{WITTEN},\cite{VENYAN}.
 
\begin{figure}[htb]
\vspace*{6.2cm}
\includegraphics{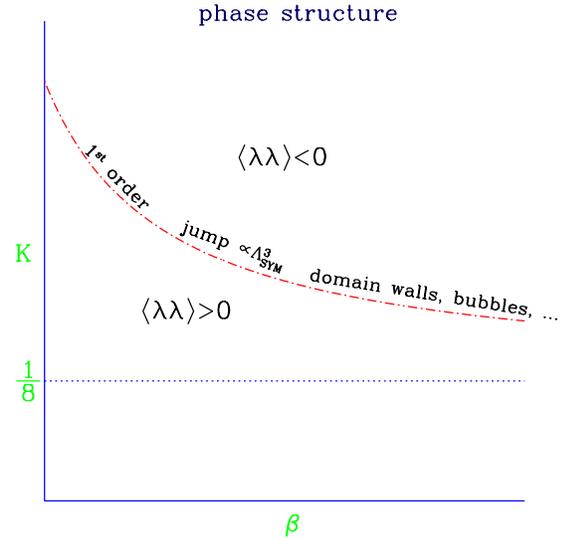}
\caption{Expected phase structure in the ($\beta,K$)-plane.
 The dashed-dotted line is a first order phase transition at zero gluino
 mass.}
\vspace*{-0.7cm}
\label{fig03}
\end{figure}

 The coexistence of several different vacua implies a first order
 phase transition at $K=K_{cr}$, at least in the continuum limit.
 At finite lattice spacing in the $(\beta,K)$-plane this might be
 represented just by cross-over lines.
 Nevertheless, the simplest possibility is a first order phase
 transition already for finite $\beta$.
 For instance, in case of $SU(2)$ gauge group the phase structure may
 be as shown by figure~\ref{fig03}.
 An alternative phase structure can be motivated by the modified
 Veneziano-Yankielowicz effective low energy action proposed recently
 by Kovner and Shifman \cite{KOVSHIF}.
 They suggest the existence of an additional massless phase with no
 chiral symmetry breaking.
 On the lattice this could lead to a more complicated structure, for
 instance, because the symmetric phase could be stable or metastable
 in some intermediate range of the hopping parameter.
%%%%%%%%%%%%%%%%%%%%%%%%%%%%%%%%%%%%%%%%%%%%%%%%%%%%%%%%%%%%%%%%%%%%%%%%
%%%%%%%%%%%%%%%%%%%%%%%%%%%%%%%%%%%%%%%%%%%%%%%%%%%%%%%%%%%%%%%%%%%%%%%%

 An interesting point is the dependence of the phase structure on the
 gauge group.
 For example, in case of $SU(3)$ there are at least three degenerate
 vacua and beyond the critical hopping parameter $K > K_{cr}$ we expect
 to be at $\Theta_{SYM}=\pi$.

 Since the Monte Carlo simulations are usually done at non-zero
 gluino mass $m_{\tilde{g}} \ne 0$, the best procedure is to derive
 analytically the dependence of different quantities on $m_{\tilde{g}}$
 \cite{GLMASS} and compare them directly with numerical data.
 In case of the gluino condensate the best possibility is to study
 the difference $\Delta\langle\lambda\lambda\rangle$ between the
 values of $\langle\lambda\lambda\rangle$ in different vacua.
 The advantage is that the additive ultraviolet divergences, due to the
 breaking of supersymmetry at finite lattice spacing, cancel.

%%%%%%%%%%%%%%%%%%%%%%%%%%%%%%%%%%%%%%%%%%%%%%%%%%%%%%%%%%%%%%%%%%%%%%%%
\subsection{Bound state spectrum}                         \label{sec3.2}

 The low energy effective actions predict that at the supersymmetric
 point the low lying states are organized in massive supermultiplets.
 The simplest possibility is to have a lowest supermultiplet generated
 by a superfield made out of gluinos \cite{VENYAN}.
 On the lattice such states are $\tilde{\pi}$, $\tilde{\sigma}$ and
 $\tilde{\chi}$ discussed already in section~\ref{sec2.3}.
 Other obvious candidates are glueball states observed in pure
 Yang-Mills theory, which might also belong to another supermultiplet.
 Of course, the notion of constituents is in general not expected to
 work perfectly in a strongly interacting theory.
 Therefore one has to be open minded towards non-trivial mixings.

\begin{figure}[htb]
\vspace*{4.0cm}
\includegraphics{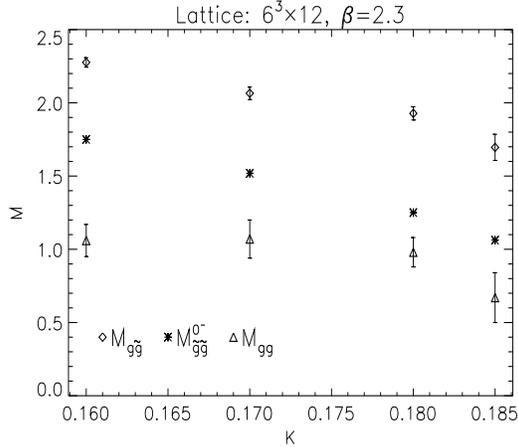}
\caption{Dependence of the lowest bound state masses 
 on the hopping parameter $K$ on $6^3 \cdot 12$ lattices at 
 $\beta=2.3$.
 Masses of states made out of gluons ($g$) and gluinos ($\tilde{g}$) are
 shown.
 (Ref.~\cite{DEMUAT})}
\vspace*{-1.0cm}
\label{fig04}
\end{figure}

 One of the main goals of the DESY-M\"unster-Athens collaboration is
 to determine the masses of low lying states in the vicinity of
 the supersymmetric point at zero gluino mass.
 First results are presented at this conference.
 In general, as shown by figure~\ref{fig04}, the supersymmetric
 degeneracy of low lying states is not yet observed.
 Further simulations closer to $K_{cr}$ and at larger $\beta$ are
 in progress.
%%%%%%%%%%%%%%%%%%%%%%%%%%%%%%%%%%%%%%%%%%%%%%%%%%%%%%%%%%%%%%%%%%%%%%%%
%%%%%%%%%%%%%%%%%%%%%%%%%%%%%%%%%%%%%%%%%%%%%%%%%%%%%%%%%%%%%%%%%%%%%%%%

%%%%%%%%%%%%%%%%%%%%%%%%%%%%%%%%%%%%%%%%%%%%%%%%%%%%%%%%%%%%%%%%%%%%%%%%
\subsection{Other questions}                              \label{sec3.3}

 Besides the phase structure and the low lying mass spectrum there are
 also other interesting questions which can be studied by Monte Carlo
 simulations.
 For instance, the direct investigation of broken supersymmetry
 Ward-Takahashi identities is possible.
 In general, the restoration of supersymmetry in the continuum limit
 has to be understood.
 Working with gluinos will teach us a lot about the simulation of
 non-even numbers of flavours and about fermions in the adjoint
 representation.
 We shall hopefully get answers to these and other related questions
 at forthcoming lattice conferences.

\raggedbottom
%%%%%%%%%%%%%%%%%%%%%%%%%%%%%%%%%%%%%%%%%%%%%%%%%%%%%%%%%%%%%%%%%%%%%%%%

\end{document}